\documentclass[smallcondensed,final]{svjour3}
\pdfoutput=1 %This is for the ArXiv submission only

\usepackage{mathbbol}
\usepackage[numbers,sort&compress]{natbib}
\usepackage{dsfont,citesort}
\usepackage{dsfont}
\usepackage{color}
\usepackage{graphicx}% Include figure files
\usepackage{bm}% bold math
\usepackage{amsmath,amsfonts}
\usepackage{amssymb}
\usepackage{stmaryrd}

\def\sing{\, {\buildrel \scriptsize{\rm (sing)} \over =}\, }

\journalname{J. Stat. Phys.}

\begin{document}

\title{
Cross-overs of Bramson's shift at the transition between pulled and  pushed fronts}
%\author{Authors}

\author{Bernard Derrida  
}

\institute{Bernard Derrida \at
              Coll\`{e}ge de France, 11 place Marcelin Berthelot, 
 75005, Paris, France \\
Laboratoire de Physique de l’Ecole Normale Sup\'{e}rieure, ENS Universit\'{e} PSL, CNRS, Sorbonne Universit\'{e}, Universit\'{e} de Paris 75005, Paris, France
\\
              bernard.derrida@college-de-france.fr
}

\maketitle

\begin{abstract}

The Bramson  logarithmic shift of the position of  pulled fronts
is a universal feature common to a large class  of monostable  traveling wave equations.  As one varies the non-linearities it so happens that one can observe, at some critical non linearity,  a transition from pulled fronts  to pushed fronts. At this transition the Bramson shift is modified. In the limit where time goes to infinity and the non-linearity  becomes critical, the position of the front exhibits  a cross-over. The goal of the present  note is  to give the expression of  this cross-over function, for a particular model which is exactly soluble, with the hope that  this expression would remain valid  for more general traveling wave equations at the transition between pulled and pushed fronts.
Other cross-over functions are also obtained,  for this particular model, to describe the dependence on initial conditions or the effect of a cut-off.

\keywords{Fisher-KPP equation, travelling waves,  Bramson shift, cut-offs}
\PACS{02.30.Jr,02.30.Mv,05.40.Jc}
\end{abstract}

\section{Introduction}
 Since its introduction  by Fisher \cite{Fisher}, and    Kolmogorov, Petrovsky, Piscounov  \cite{KPP} 
 the F-KPP equation 
\begin{equation}
{d u \over dt} =\frac{d^2 u}{d x^2} + u (1-u)
\label{AX-1}
\end{equation}
has  played a central role in the theory of partial differential equations \cite{Aronson,HR} and of the Branching Brownian motion \cite{McKean} (see also \cite{CR,DMS,DS1,bd}).
When the  initial condition  $u(x,0) \equiv u_0(x)$ is a step function
\begin{equation}
u_0(x) =  \left\{\begin{array}{lll} 1 & \text{   for   } & x \le 0 \\
0 &  & x > 0 \end{array} \right.
\label{AX-2}
\end{equation}
or  when it decays fast enough  as for example 
\begin{equation}
u_0(x) =  \left\{\begin{array}{lll} 1 & \text{   for   } & x \le 0 \\
 \exp(- s \,  x) 
  &  & x > 0 \end{array} \right.
\ \ \ \ \ \ \ \text{with} \ \  s >1
\label{AX-3}
\end{equation}
it is known since the work of Bramson \cite{Bramson1,Bramson2,Hamel,AS} that in the long time limit the solution  becomes  a front 
\
\begin{equation}
u(x,t) \simeq W(x- \mu_t)
\label{AX-4}
\end{equation}
 located at position 
\begin{equation}
\mu_t \simeq  2 t - \frac{3}{2} \ln t + A\big(\{u_0(x)\}\big)  + o(1) \ . 
\label{AX-5}
\end{equation}
The traveling wave  shape $W$ in (\ref{AX-4}) is a solution of an ordinary non-linear differential equation
\begin{equation}
2  W'(z) +W''(z) + W(z)(1-W(z))=0  
\label{AX-6}
\end{equation}
which satifies  $ W(-\infty)=1 \  ; \   W(\infty)=0$
and the shift $A(\{u_0(x)\})$ depends on the initial condition $u(x,0)$ (and also on the choice made to fix  the solution of (\ref{AX-6}), for example $W(0)=\frac{1}{2}$).
What is remarkable is that the $\log t$ term in (\ref{AX-5})  does not depend on the initial condition \cite{Bramson1,Bramson2} (as long as it decays fast enough). In fact this  Bramson shift $-\frac{3}{2} \log t$ remains the same for all monostable traveling wave equations of the form 
\begin{equation}
{d u \over dt} =\frac{d^2 u}{d x^2} + g(u)
\label{AX-7}
\end{equation}
when the non-linearity $g(u)$ satisfies
\begin{equation}
\label{AX-8}
g(0)=0  \ \ \ \ \ ; \ \ \ \ g'(0)=1 \ \ \ \ ; \ \ \ \ g(1) =0  \ \ \ \  ;
\end{equation}
and 
\begin{equation}
 0 < g(u) < u \ \ \ \text{for} \ \ \     0 < u < 1
\label{AX-9}
\end{equation}

When the  position of the traveling wave is  given by (\ref{AX-5}) one says that the front is pulled \cite{Saa}. Thus   when the non-linearity satisfies condition (\ref{AX-9}), one is in the case of "pulled fronts".
When (\ref{AX-9}) is not satisfied, keeping $g(u)>0$ in $(0,1)$,  but  with  some values of $u$ where $g(u)>u$, the long time asymptotics may still given by (\ref{AX-4},\ref{AX-5}) for steep enough initial conditions (\ref{AX-2},\ref{AX-3})  (in which  case the front is still  pulled) or may lead to a different long time asymptotics (called "pushed fronts")
with  the position $\mu_t$ of the front becoming
\begin{equation}
\mu_t = v t + C\big(\{u_0(x)\}\big) + o(1) \ \ \ \ \ \text{with} \ \ \ \ \ v>2
\label{AX-10}
\end{equation}
	(for a general non-linearity $g(u)$ as   in (\ref{AX-7})  the shape $W$ of the wave front is  solution of $v W' + W'' + g(W)= 0$  with  $v=2$ in the pulled case and some  $v>2$ in the pushed case).

For example it is well known  \cite{GGHR} that for
\begin{equation}
g(u)=
u (1-u) (1+ 2a u)
\label{AX-11}
\end{equation}
the non linearity satisfies the condition (\ref{AX-9}) for $a\le \frac{1}{2}$ but the front remains pulled for $a<1$. For $a >1$ the front becomes pushed and, for steep enough initial conditions, the asymptotic velocity $v = 
\sqrt{a }+ 
\sqrt{\frac{1}{a} }
$
 with, for $a>1$,  a  front shape given by  $W(x)= \Big(1 + \exp[\sqrt{a}\  x] \Big)^{-1} $.
(Other examples can be found in \cite{ahr2}).

At the transition between the pulled case  and  the pushed case  \cite{ahr1,ahr2,AHS} (i.e. for $a=1$ for the example (\ref{AX-11})), the long time asymptotics of  $\mu_t$ is modified to become
\begin{equation}
\mu_t \simeq  2 t - \frac{1}{2} \ln t + B\big(\{u_0(x)\}\big)  + o(1)
\label{AX-12}
\end{equation}
where, 
 the logarithmic term is now $-\frac{1}{2} \log t$
 instead of  $-\frac{3}{2} \log t$
 in (\ref{AX-5}).

One goal of the present work is to try to understand the cross-over between the three different asymptotics (\ref{AX-5},\ref{AX-10},\ref{AX-12}). Let us imagine that the non-linearity $g_a(u)$ in (\ref{AX-7}) depends on a parameter $a$ and that there is a critical value $a_c$ of this parameter which separates the pulled case for $a < a_c$ from  the pushed case for $a > a_c$.  (This critical value $a_c=1$ for the example (\ref{AX-11})).  Already one could guess that, in (\ref{AX-5}) and in (\ref{AX-10}), the constants 
$A\big(\{u_0(x)\}\big)  \to \infty$ as $a \to a_c^-$ and  
$C\big(\{u_0(x)\}\big)  \to -\infty$ as $a \to a_c^+$.
 But what to expect for $\mu_t$  in the regime where $|a-a_c|$ is very small and $t$ is very large?

We are going to obtain   the asymptotics of $\mu_t$ in this cross-over regime for a system \cite{bbd}   which can be viewed as the hydrodynamic limit of the $N$-BBM or of the $L$-BBM \cite{BD,BDMM3,M1,Pain,DS,DR,Carinci}. In this simple model \cite{bbd}
 the evolution of $u$ is given by
\begin{align}
& \frac{du}{dt}= \frac{d^2 u }{dx^2} + u  \ \ \ \ \text{for} \ \ \ \ x > \mu_t
\nonumber
\\
& u(\mu_t,t) = 1
\label{AX-13}
\\
& \partial_x u(\mu_t,t)=-a   \ \ \ \ \text{with} \ \ \ \  a>0
\nonumber
\end{align}
% \begin{equation}
% \left\{
% \begin{array}{l}
%  \frac{du}{dt}= \frac{d^2 u }{dx^2} + u  \ \ \ \ \text{for} \ \ \ \ x > \mu_t
% \\
%  u(\mu_t,t) = 1 \\
% \partial_x u(\mu_t,t)=-a
% \end{array}
% \right.
% \label{AX-13}
% \end{equation}
This is a free boundary problem  \cite{Carinci,HH}   and, given the initial condition $u_0(x)$,  the evolution (\ref{AX-13}) determines both $u(x,t)$ 
and $\mu_t$. (To avoid complicated discussions,  we consider here only decreasing initial conditions  such that $\mu_0=0, u_0(0)=1$ and $u(\infty)=0$).
This  problem was studied in \cite{bbd}, the parameters $\alpha$ and $\beta$ in \cite{bbd} taking here the values  $\alpha=1$ and $\beta=-a$. 
 In this simple model,  as in its earlier lattice version \cite{BD1}, the non-linearity comes only from the  boundary condition at $ x=\mu_t$ (see (\ref{AX-13}) which can be interpreted as    $\frac{du}{dt}=0 $ for $ u \ge 1$). The reason why explicit calculations are possible for this example is that there exists an exact relation which was established in \cite{bbd} (see Equation (6) of \cite{bbd}) between the position $\mu_t$ and the initial condition $u_0(x)$
which can be written here as 
\begin{equation}
1 + r \int_0^\infty dz \ u_0(z) \ e^{r z}=  (1-a r) \int_0^\infty dt \,  e^{r \mu_t-(1+r^2)t}
\label{AX-14}
\end{equation}
(This relation remains valid when $r$ varies  as long as the integrals on both sides of (\ref{AX-14}) converge).

Based on this equation it
 has been shown  (see Equations (36,39,40) of \cite{bbd}) that for a step initial condition (\ref{AX-2}) (as well as for steep enough initial conditions as in (\ref{AX-3})) 

% \begin{align}
% \nonumber
% \mu_t=& 2 t -{3 \over 2} \log t + A\Big(\{u_0(x)\}\Big)  -  {3 \sqrt{\pi} \over \sqrt{t}} +  o\left({ 1\over \sqrt{t}} \right)
 % &  \  \text{ for}\ \   & a <1  
% \\
% \label{AX-15}
% \mu_t=& 2 t -{1 \over 2} \log t + B\Big(\{u_0(x)\}\Big)  -  { \sqrt{\pi} \over 2\sqrt{t}} +  o\left({ 1\over \sqrt{t}} \right)
 % &  \  
% \text{ for}
% \ \   & a =1  
% \\ 
% \mu_t=& \left({2+(a-1)^2 \over a}\right) t + 
%  C\Big(\{u_0(x)\}\Big) 
%   + o(1)
%  & \ 
% \text{ for}
% \ \   & a =1  
% \nonumber
% \end{align}

\begin{align}
\nonumber
\text{ For}\ \   & a <1  & \ \ \ &
\mu_t= 2 t -{3 \over 2} \log t + A\Big(\{u_0(x)\}\Big)  -  {3 \sqrt{\pi} \over \sqrt{t}} +  o\left({ 1\over \sqrt{t}} \right)
\\
\label{AX-15}
 & a  =1 & &
\mu_t= 2 t -{1 \over 2} \log t + B\Big(\{u_0(x)\}\Big)  -  { \sqrt{\pi} \over 2\sqrt{t}} +  o\left({ 1\over \sqrt{t}} \right)
\\ 
& a>1 & & 
\mu_t= \left({2+(a-1)^2 \over a}\right) t + 
 C\Big(\{u_0(x)\}\Big) 
  + o(1)
\nonumber
\end{align}
where $A,B,C$   are  time-independent constants whose expressions are not known and depend on the initial condition $u_0$.
\ \\ \ \\ 

What we are going to show in section \ref{S2} is that, in the scaling regime where
\begin{equation}
a-1= \epsilon \ll 1 \ \ \ \ \ \text{and} \ \ \ \ \ t= O(\epsilon^{-2}) \ \ \ , 
\label{AX-16}
\end{equation}

\begin{equation}
\mu_t =2 t -\frac{1}{2} \log t + \Psi_1( \epsilon \sqrt{t})  + 
 B\big(\{u_0(x)\}\big)+  o(1)
\label{AX-17}
\end{equation}
where
\begin{equation}
\label{AX-18}
 \Psi_1(z)=\log \left[1   + 2 z   e^{z^2 } \int_{-\infty}^z dv \  e^{-v^2}   \right] 
\end{equation}
and
\begin{equation}
 B\big(\{u_0(x)\}\big) =  \log \left[1 + \int_0^\infty  dz \ e^z \ u_0(z) \right]
- \frac{1}{2} \log \pi 
\label{AX-19}
\end{equation}
We will see that  the limits $\epsilon \sqrt{t} \to \pm \infty$  are consistent with (\ref{AX-15}).

In section \ref{S3} we will study the cross-over due to initial conditions, i.e. when the decay rate in  $s \to 1 $ in (\ref{AX-1}). We will discuss separately  the pulled case (i.e. when $a<1$ in (\ref{AX-13})) and   the transition point $a=1$ beween the pushed and the pulled cases.

Lastly, in section \ref{S4}, we will discuss the effect of a cut-off on the velocity of the traveling wave (where one imposes the additional boundary condition that $u(\mu_t + L)=0$). In this case too, there  a crossover  near the transition between the pulled and the pushed cases when the cut-off parameter $L \to \infty$ at the same time as the distance to the transition point $\epsilon= a-1 \to 0$, keeping the  product $L \epsilon = O(1)$.

\section{Cross-over of the Bramson shift at the transition between pushed and pulled fronts}
\label{S2}
All our analysis is based on the exact relation (\ref{AX-14}) (which was established in \cite{bbd})) between the initial condition $u_0(x)$ and the position of the front $\mu_t$.
Assuming that $a=1 + \epsilon$  as in (\ref{AX-13}) and (\ref{AX-16}) one can rewrite (\ref{AX-14}) as

\begin{equation}
\left[1 + r \int_0^\infty dz \ u_0(z)\  e^{r z} \right]
\ \frac{1}{ 1 - r - \epsilon r}
=   
\int_0^\infty dt \,  e^{r \mu_t-(1+r^2)t}
\label{BX-1}
\end{equation}

As in \cite{bbd} the main idea is to equate the singular parts of the two sides of (\ref{BX-1})  when $r \to 1$.
 If one expands 
the left hand side
in powers of $\epsilon$, one gets, by keeping  at each order in $\epsilon$  the most singular term at $r=1$,
one gets 

\begin{equation}
\left[1 + r \int_0^\infty dz\  u_0(z)\  e^{r z} \right]
\frac{1}{ 1 - r - \epsilon r}
 \sing  
\left[1 +  \int_0^\infty dz \ u_0(z)\  e^{ z} \right] 
\sum_{n=0}^\infty \frac{\epsilon^n}{(1-r)^{n+1}}
\label{BX-2}
\end{equation}
(where $\sing$ means that,  at each order in $\epsilon$, the most singular terms in the limit $r \to 1$  are the same on both sides of $\sing$.)
Then using the fact  that
\begin{equation}
{1 \over (1-r)^{n+1}}= {1 \over \Gamma({n+1 \over 2}) } \int_0^\infty dt \ t^{n-1 \over 2} e^{- (1-r)^2 \, t }
\label{BX-3}
\end{equation}
one gets for the left-hand side of (\ref{BX-2})
\begin{align}
\left[1 + r \int_0^\infty dz \ u_0(z)  \ e^{r z} \right] &
\ \frac{1}{ 1 - r - \epsilon r}
 \label{BX-4}
 \ \sing   \\ &
\left[1 + \int_0^\infty dz \ u_0(z)  \ e^{ z} \right] 
\int_0^\infty dt 
\frac{ G_1( \epsilon \sqrt{t}) }{\sqrt{t}} 
\  e^{-(1-r)^2t}
\nonumber
\end{align}
where $G_1(z)$ is defined by
\begin{equation}
 G_1(z)= \sum_{n \ge 0} \frac{z^n}{ \Gamma({n+1 \over 2})}
 \label{BX-5}
\end{equation}

On the other hand, as far as the singular part at $r=1$ is concerned, one could replace the lower bound in the integral on   the right hand side of (\ref{BX-1}) by any time $t_0$
\begin{equation}
 \int_0^\infty dt \,  e^{r \mu_t-(1+r^2)t}
\sing
 \int_{t_0}^\infty dt \,  e^{r \mu_t-(1+r^2)t}
 \label{BX-6}
\end{equation}
 (since the integral from $0$ to $t_0$ is an analytic function of $r$). Therefore, as $t_0$ can be arbitrary large, it is clear that the  singular part at $r=1$ of the right hand side of (\ref{BX-1}) is fully determined by  the long time asymptotics of $\mu_t$.  
One has for the right hand side of (\ref{BX-1})
\begin{align}
   \int_0^\infty dt \,  e^{r (\mu_t-2 t) -(1-r)^2 t}
 &  \sing 
    \int_0^\infty dt \,  e^{ (\mu_t-2 t) -(1-r)^2t} \left[ 1 + (r-1) (\mu_t-2 t) + \cdots \right]
\nonumber
 \\  & \sing
    \int_0^\infty dt \,  e^{ (\mu_t-2 t) -(1-r)^2t} 
 \label{BX-7}
\end{align}
where, again, in the last line,  we keep only the first term because we look for the most singular term in $r$ at each order in powers of $\epsilon$.

Equating the left hand side (\ref{BX-4})  and the right hand side of (\ref{BX-7})  of (\ref{BX-1}) for a whole neighborhood of $r=1$ leads to  
the following large $t$ asymptotics for $\mu_t$
\begin{equation}
 e^{ \mu_t-2 t }
\simeq 
\left[1 +  \int_0^\infty dz \ u_0(z) \ e^{ z} \right] 
\frac{ G_1( \epsilon \sqrt{t}) }{\sqrt{t}} 
\label{BX-8}
\end{equation}
which  can be rewritten as in (\ref{AX-17},\ref{AX-18},\ref{AX-19}) using the expression (\ref{FX-7}) of $G_1(z)$ established in the Appendix A.
One can also note that  the $z \to \pm \infty$ asymptotics (\ref{FX-8}) of the function $G_1$ allow to recover both  the pulled and the pushed regimes (\ref{AX-15}).

As explained in Appendix B, by analyzing the next singular terms at $r=1$ in the equation  (\ref{BX-1}), one can get cross-over functions of the next terms in the asymptotics of $\mu_t$, i.e. vanishing corrections to (\ref{AX-17}) in the long time limit 
\cite{ES,bbd,Henderson,BBHR,NRR,MM,Graham,ahr1,ahr2}.

\section{Cross-over due to initial conditions}
\label{S3}
In this section we analyze the cross-over due to the initial condition both in the pulled case and at the boundary between the pulled and the pushed case. As we will see the analysis is very similar to what was done in section \ref{S2}.

\begin{enumerate}
\item {\it The pulled case ($a<10$ and $s \to 1$):}

In the pulled case, (i.e. for $a<1$ in the problem (\ref{AX-13})), we want to consider the cross-over regime where  $s \to 1$ in the initial condition (\ref{AX-3}) and $t \to \infty$.
If we write
\begin{equation}
s=1 + \varphi \ \ \ \ \ \text{with} \ \ \ \varphi \ll 1
\label{CX-1}
\end{equation}
the relation  (\ref{AX-14}) between the initial condition (\ref{AX-3}) and the position of the front $\mu_t$ becomes
\begin{equation}
\frac{ 1 + \varphi}{(1-a r) \ (1+\varphi -r)} =   \int_0^\infty dt \,  e^{r \mu_t-(1+r^2)t}
\label{CX-2}
\end{equation}
Now if we look,  at each order in powers of $\varphi$, for  the most singular term at $r=1$ of the left hand side of (\ref{CX-2})
one gets 
\begin{equation}
\frac{ 1 + \varphi}{(1-a r) \ (1+\varphi -r)}    \sing
\frac{ 1 }{(1-a ) \ (1+\varphi -r)}  
\label{CX-3}
\end{equation}
Copying what was done in (\ref{BX-2},\ref{BX-3},\ref{BX-4}) one gets for the left hand side of (\ref{CX-2})
\begin{equation}
\frac{ 1 + \varphi}{(1-a r) \ (1+\varphi -r)}  \sing
\frac{ 1} {(1-a ) 
\ (1+\varphi -r)}  \sing
\frac{ 1} {1-a } 
\int_0^\infty dt 
\frac{ G_1( -\varphi \sqrt{t}) }{\sqrt{t}} 
\  e^{-(1-r)^2t}
\label{CX-4}
\end{equation}
The analysis of the right hand side of (\ref{CX-2}) is exactly the same  (see (\ref{BX-7})) as in section \ref{S2}. Therefore one gets for the pulled case
\begin{equation}
 e^{ \mu_t-2 t }
\simeq
\left[\frac{ 1} {1-a } \right] 
 \ \frac{ G_1( -\varphi \sqrt{t}) }{\sqrt{t}} 
\label{CX-5}
\end{equation}
so that, using the expression (\ref{FX-7}) one obtains
\begin{equation}
\mu_t =2 t -\frac{1}{2} \log t - \log(1-a) + \Psi_1( -\varphi \sqrt{t})   
 +  o(1)
\label{CX-6}
\end{equation}
where the function $\Psi_1 $  is given in (\ref{AX-18}).

\ \\ 
\item {\it The boundary between the pulled and the pushed  cases  ($a=1$ and $s \to 1$):}

Let us now look at the case $a=1$ in (\ref{AX-13}), i.e. at the transition between the pushed and the pulled case. The left hand side of (\ref{BX-1}) becomes 
\begin{equation}
{1 + \varphi \over (1-r) (1+\varphi -r)} = \sing \sum_{n \ge 0} (-1)^n {\varphi^n\over (1-r)^{n+2}}= 
\int_0^\infty dt 
 G_2( -\varphi \sqrt{t}) 
\  e^{-(1-r)^2t}
\label{CX-7}
\end{equation}
where we have used (\ref{BX-4}) and 
the function $G_2$ is defined by
\begin{equation}
 G_2(z)= \sum_{n \ge 0} \frac{z^n}{ \Gamma({n+1 \over 2})}
 \label{CX-8}
\end{equation}
The expression of the most singular terms in the right hand side of (\ref{BX-1}) remains given by (\ref{BX-7}), so that, equating the most singular parts (\ref{CX-7}) and (\ref{BX-7}) of the two sides of  (\ref{BX-1})
one gets 
\begin{equation}
\mu_t=2 t  + \Psi_2(-\sqrt{\varphi} t) 
 \label{CX-9}
\end{equation}
where the function $\Psi_2$ is given by (see (\ref{FX-9})
\begin{equation}
\Psi_2(z)=  \log \left[ 
 {2\over \sqrt{\pi}}    e^{z^2} \int_{-\infty}^z e^{-t^2} dt \right]
 \label{CX-10}
\end{equation}

\end{enumerate}

\section{The effect of a cut-off  on the velocity of the traveling wave} 
\label{S4}
For the F-KPP equation it is well established that the first correction to the velocity of the traveling wave due to a weak white noise \cite{MulS,MMQ,BG} is the same as the correction  due to a cut-off \cite{BD,DPK,BDL}.
In this section we  calculate this correction  due to a cut-off for the model (\ref{AX-13}). We will analyze successively the pulled case, the  pushed case and   the cross-over near the transition between pulled and pushed fronts.

A traveling wave  in the problem (\ref{AX-13})  moving at velocity $v$ is a solution $u$ of the form  
\begin{equation}
u(x,t)=W(x-v t)=\left\{\begin{array}{lll} {\cal A} \exp\left[{-\gamma(x-v t)}\right]  + {\cal B}\exp\left[ -\gamma^{-1} (x-v t)  \right]  & \text{for} & x  \ge v t
\\
1  & \text{for} & x \le  v t
\end{array} \right.
\label{DX-1}
\end{equation}
where  the amplitudes ${\cal A}$  and ${\cal B}$ satisfy 
\begin{equation}
W(0)= {\cal A}+{\cal B} = 1
\ \ \ \ \ ; \ \ \ \ \ 
W'(0)=-{\cal A }\gamma-\frac{{\cal B}}{\gamma}=-a \ \ \ .
\label{DX-2}
\end{equation}
These equations (\ref{DX-2})   determine ${\cal A}$ and ${\cal B}$ in terms of $\gamma$
with a    velocity $v$   and a front position $\mu_t$ given  by 
\begin{equation}
v = \gamma + \frac{1}{\gamma}
\ \ \ \ \ ; \ \ \ \ \ 
\mu_t  = v \, t 
\label{DX-3}
\end{equation}

% \\
If one introduces  a cut-off by requiring that the front vanishes at position $x =v t + L$, one gets an additional equation which fixes $\gamma $ and therefore the velocity $v$
\begin{equation}
W(L)= {\cal A} \exp[-\gamma L] + {\cal B} \exp\left[-\gamma^{-1} L  \right]   =0 \ \ \ .
\label{DX-4}
\end{equation}
% \\
For
\begin{equation}
a=1+\epsilon
\label{DX-5}
\end{equation}
 $\gamma$ should therefore satisfy
\begin{equation}
\exp\left(L \frac{1-\gamma^2}{\gamma} \right) = -\gamma \  \frac{1+ \epsilon-\gamma}{1-\gamma-\epsilon \gamma}  \ . 
\label{DX-6}
\end{equation}
To determine 
  the large $L$ correction to the velocity $v$ one simply has to calculate the  large $L$ dependence of the solution $\gamma $ of (\ref{DX-6}).  One can distinguish the three following situations:

\begin{enumerate}
\item {\it The pulled case} ($\epsilon<0$):

In absence of a cut-off, i.e. in the limit  $L \to \infty$, the velocity $v \to 2$ and $\gamma \to 1$.
For large $L$, the solution of (\ref{DX-6}) is
\begin{equation}
\gamma= 1 \pm i \frac{\pi}{L} + \cdots
\end{equation}
 and one recovers the expected correction to the  velocity
for pulled fronts \cite{BD,DPK,BDL}
\begin{equation}
v= 2- \frac{\pi^2} {L^2} + \cdots \ \ \ .
\label{DX-7}
\end{equation}

\item {\it The pushed case} $(\epsilon>0$):

In the pushed case, ${\cal A}$ in (\ref{DX-1}) vanishes  when there is no cut-off, i.e. for $L \to \infty$, so that $\gamma={1 \over a}={1 \over 1 + \epsilon}$.  For large $L$, the solution of (\ref{DX-6}) is
\begin{equation}
\gamma=\frac{1}{1 + \epsilon} + \frac{2 \epsilon + \epsilon^2}{(1+\epsilon)^3}
\exp\left[- \frac{2 \epsilon + \epsilon^2}{1 + \epsilon} L \right] + \cdots
\end{equation}
 and the correction to the velocity  is exponentially small in $L$
\begin{equation}
v\simeq2 + \frac{\epsilon^2 }{1+\epsilon} -
\frac{(2 \epsilon+\epsilon^2)^2 }{(1+\epsilon)^3} \exp\left[- \frac{2 \epsilon + \epsilon^2}{1 + \epsilon} L \right] + \cdots  
\label{DX-8}
\end{equation}
as expected in the pushed case \cite{KNS}.
\ \\ 

\item {\it The cross-over  between the pulled and the pushed case}:
% (large $L$ and $L \epsilon=O(1)$)

For $\epsilon$ small and large $L$, keeping the product $ L \epsilon$ of order 1, if one writes
\begin{equation}
\gamma \simeq  1 +i {\chi \over L} 
\end{equation}
and solves  (\ref{DX-6}) in the large $L$ limit,    $\chi$  is  solution of 
\begin{equation}
\chi = (L \epsilon) \, \tan(\chi)
\label{DX-9}
\end{equation}
(one should choose the solution such that $\chi \to \frac{\pi}{2} $ as $L \epsilon \to 0$: for other choices like  $\frac{(2 n +1)\pi}{2}$ with $n=1,2,3  \cdots$ the traveling wave $W(x)$ would not remain  positive in the interval $(0,L)$).
Therefore 
\begin{equation}
v=  2 -\frac{\chi^2}{L^2} + \cdots 
\label{DX-10}
\end{equation}
where $\chi$ is the  function of $L \epsilon$, solution of (\ref{DX-9}).
For example for $L \epsilon$ small one gets  

\begin{equation}
\chi= 
\frac{\pi}{2 } 
- \frac{2}{\pi} L \epsilon  
- \frac{8}{\pi^3} (L \epsilon)^2  
+ \left(\frac{8 }{3 \pi^3} - \frac{64}{\pi^5} \right) (L \epsilon)^3   + O \Big((L \epsilon)^4\Big)
\end{equation}
giving
\begin{equation}
v=  2   + \frac{1}{L^2} \left[  
 -\frac{\pi^2}{4 } + 2 (L \epsilon)    + \frac{4}{\pi^2} (L \epsilon)^2 +\left(\frac{8}{3 \pi^2}-\frac{32}{\pi^4}\right)  (L \epsilon)^3  +   O\Big((L \epsilon)^4\Big) \right]
\ . 
\label{DX-11}
\end{equation}

In the particular case where $\epsilon=0$, i.e. right  at the transition between the pulled and the pushed case, $\chi= \frac{\pi}{2}$ and the correction to the velocity becomes 
\begin{equation}
v \simeq 2 - \frac{\pi^2}{4 L^2}
\label{DX-12}
\end{equation}
in constrast to the expression (\ref{DX-7}) for the pulled  case.

For $L \epsilon \to -\infty$, 
one also gets
\begin{equation}
v=2-\frac{\pi^2}{L^2} \left(1 + \frac{2}{L \epsilon} + \frac{3}{L^2 \epsilon^2} + \cdots \right) 
\end{equation}
which matches with (\ref{DX-7}).

As  the product $L \epsilon $ varies from $-\infty$ to $ 1$ the solution $\chi$ of (\ref{DX-9}) decreases  from $\pi$ to $=0$ (taking the value $\frac{\pi}{2}$ when $L \epsilon=0$).
For $ L \epsilon > 1$, $\chi$  solution of (\ref{DX-9}) becomes imaginary: $\chi=i \chi'$
and the velocity becomes
\begin{equation}
v=2 + \frac{\chi'^2}{L^2} + \cdots
\label{DX-13}
\end{equation}
where $\chi'$ is solution of 
\begin{equation}
\chi'=  L \epsilon \tanh \chi' \ \ \ \ \ \text{with} \ \ \ \ \ 0< \chi'  
\end{equation}
(There is no singularity at $L \epsilon=1$. The solutions (\ref{DX-11}) and (\ref{DX-13}) for $L \epsilon < 1$ and $L \epsilon >1$ are analytic continuations of each other.)

For $L \epsilon \to + \infty$ \large, one has
\begin{equation}
\chi'   =    L \epsilon  (1 - 2 e^{-L \epsilon} + \cdots)  \end{equation}
\begin{equation}
v   =    2 + \epsilon^2  -
4 \epsilon^2  e^{- 2 L \epsilon  } + \cdots
\end{equation}
which also matches with (\ref{DX-9}).
\end{enumerate}

\section{Conclusion}
The goal of the present  work was to derive explicit expressions (\ref{AX-17},\ref{AX-18},\ref{AX-19}) of the cross-over functions describing  the position of the front near the transition between pulled and pushed fronts for 
the model (\ref{AX-13}).
For the same model, other cross-over functions have been obtained for the dependences on initial conditions (\ref{CX-6},\ref{CX-9},\ref{CX-10}) or for the shift of velocity (\ref{DX-9},\ref{DX-10},\ref{DX-12},\ref{DX-13}) due to a cut-off.

All these cross-over functions were obtained for the very specific model (\ref{AX-13}).
Of course it would be  interesting to know their degree of universality,  in particular if they  could also describe the same regimes for other non-linear traveling wave equations, such as those discussed in \cite{ahr1,ahr2}.

Another interesting question would be to understand the effect of a weak noise on the velocity of traveling waves.
 For pulled fronts it is well established \cite{MMQ,BG} that the leading correction  (\ref{DX-10}) to the velocity due to a weak noise can be predicted by a cut-off  argument \cite{BD}.
Does the same cut-off argument   give  the correct correction (\ref{DX-12}) due a weak noise at the transition between  pulled and pushed front and even in the whole cross-over regime?

\ \\
{\bf Acknowledgements:}
\\  I would like to thank the Centre de Recherches Math\'{e}matiques de l'Universit\'{e} de Montr\'{e}al, for its hospitality. This work was started there during the program {\it Probability and PDEs}. At the origin of the present work are the lectures 
on pushmi-pullyu fronts
that L. Ryzhik  gave during this program. I would like to thank him for several stimulating discussions.

%\appendix
\section*{Appendix A}
\renewcommand{\theequation}{A.\arabic{equation}}
% reset the counter
\setcounter{equation}{0}

\label{S5}
In this appendix we obtain explicit expressions of the sums of the series which appear in (\ref{BX-3},\ref{CX-5},\ref{CX-8}).  We are going to  show that
the function $G_\alpha(z)$ defined by 
\begin{equation}
G_\alpha(z) = \sum_{n \ge 0} {z^n \over \Gamma(\frac{n +\alpha}{2} )} 
\label{FX-1} 
\end{equation}
can be written, for $\alpha > 2$,  in terms of an incomplete Gamma function as
\begin{equation}
G_\alpha(z)=  e^{z^2} z^{2-\alpha} \int_0^{z^2} e^{-y}
dy
\left[ {y^{\alpha-4 \over 2} \over \Gamma({\alpha-2 \over 2})}
+  {y^{\alpha-3 \over 2} \over \Gamma({\alpha-1 \over 2})} \right]
\label{FX-2} 
\end{equation}
To do so, one can check that the series in (\ref{FX-1})
satisfies 
\begin{equation}
{d G_\alpha \over dz} = 2 z G_\alpha- {\alpha-2 \over z} G_\alpha 
+ {2 \over z \, \Gamma({\alpha-2 \over 2})}
+ {2 \over  \Gamma({\alpha-1 \over 2})}
\label{FX-3} 
\end{equation}
and one gets for $\alpha > 2$
\begin{equation}
G_\alpha(z)= 2 e^{z^2} z^{2-\alpha} \int_0^z e^{-u^2} 
du
\left[ {u^{\alpha-3} \over \Gamma({\alpha-2 \over 2})}
+  {u^{\alpha-2} \over \Gamma({\alpha-1 \over 2})} \right]
\label{FX-4} 
\end{equation}
which, is equivalent to (\ref{FX-2}).

Moreover it is also easy to check that for,  any $\alpha$,
the sum in (\ref{FX-1})   satisfies
\begin{equation}
G_\alpha(z) = \frac{1}{\Gamma(\frac{\alpha}{2})} + z \ G_{\alpha +1}(z)
\label{FX-5} 
\end{equation}
Therefore (\ref{FX-2}) together with (\ref{FX-5}) gives a closed expression of $G_\alpha$ for any $\alpha$. 

Using the explicit expression (\ref{FX-2}) or the differential equation (\ref{FX-3}) it is easy to extract the  large $z$ asymptotic series of $G_\alpha$:

\begin{align}
G_\alpha (z) \simeq & 2 z^{2-\alpha} e^{z^2}- \sum_{n \ge 1}{1 \over z^n \, \Gamma(\frac{\alpha-n}{2})}  \ \ \ \ \
 & \text{as} \ \ & z\to + \infty  \nonumber \\
&    \label{FX-6} \\ 
G_\alpha (z) \simeq & - \sum_{n \ge 1}{1 \over z^n \, \Gamma(\frac{\alpha-n}{2})}  \ \ \ \ \
 & \text{as} \ \ & z\to -\infty \nonumber
\end{align}

In the particular cases $\alpha=1$  and $\alpha=2$, one then gets from (\ref{FX-2},\ref{FX-6})
\begin{equation}
G_1(z)= {1\over \sqrt{\pi}} +{2\over \sqrt{\pi}}   z e^{z^2} \int_{-\infty}^z e^{-t^2} dt 
\label{FX-7}
\end{equation}
with the following asymptotics
\begin{align}
\nonumber
\text{For} \ \ & z  \to  - \infty &  & G_1(z)  \simeq     \frac{1}{\sqrt{\pi} \, z^2} + O \left(\frac{1}{z^4} \right)
\\
\label{FX-8}
\text{For} \ \ &  z \to 0 &  &   G_1(z)  =  {1 \over  \sqrt{\pi  } } + z + O(z^2) 
\\
\nonumber
\text{For} \ \ & z  \to + \infty &  & G_1(z)  \simeq    2 \, z \,  e^{z^2 } + \frac{1}{\sqrt{\pi} \, z^2} + O \left(\frac{1}{z^4} \right)
\end{align}
and
\begin{equation}
G_2(z)= {2\over \sqrt{\pi}}    e^{z^2} \int_{-\infty}^z e^{-t^2} dt 
\label{FX-9}
\end{equation}
with the following asymptotics
\begin{align}
\nonumber
\text{For} \ \ & z  \to  - \infty &  & G_2(z)  \simeq   -  \frac{1}{\sqrt{\pi} \, z} + O \left(\frac{1}{z^3} \right)
\\
\label{FX-10}
\text{For} \ \ &  z \to 0 &  &   G_2(z)  = 1+  {2 \over  \sqrt{\pi  } }  z + O(z^2) 
\\
\nonumber
\text{For} \ \ & z  \to + \infty &  & G_2(z)  \simeq    2 \,  e^{z^2 } - \frac{1}{\sqrt{\pi} \, z} + O \left(\frac{1}{z^3} \right)
\end{align}
\
\section*{Appendix B}
\renewcommand{\theequation}{B.\arabic{equation}}
% reset the counter
\setcounter{equation}{0}
\label{S6}
The goal of  this appendix  is to push further the analysis  of the most singular terms  at $r=1$ of the  equality (\ref{BX-1}), by looking now at the two most singular terms in order to study vanishing corrections to the front position \cite{ES,bbd,Henderson,BBHR,NRR,MM,ahr1,ahr2}.
To simplify the discussion we will  only consider  the case of a step initial condtion, i.e. to the case where $u_0(z)=0$ for $z>0$.
Looking,
at each order in $\epsilon$, 
 for the two most singular terms 
in  the left hand side of (\ref{BX-1}) 
one gets
\begin{equation}
\left[1 + r \int_0^\infty dz \ u_0(z)\  e^{r z} \right]
\ \frac{1}{ 1 - r - \epsilon r} \sing  
\sum_{n\ge 0}  \epsilon^n \left[\frac{1}{(1-r)^{n+1}}-  \frac{n}{(1-r)^n} \right]
\label{GX-1}
\end{equation}
Let us now assume,   for large $t$ and $\epsilon \sqrt{t} = O(1)$, the following generalization of (\ref{BX-8})
\begin{equation}
e^{\mu_t-2 t} = \frac{1}{\sqrt{t} }
 \sum_{n\ge 0} b_n (\epsilon \sqrt{t})^n + 
 \frac{\log t }{t}
 \sum_{n\ge 1} c_n (\epsilon \sqrt{t})^n + 
 \frac{1 }{t}
 \sum_{n\ge 0} d_n (\epsilon \sqrt{t})^n 
+ o\left(\frac{1} {t} \right)
\label{GX-3}
\end{equation}
 Using the identity (\ref{BX-3}) 
\begin{equation}
 \int_{t_0}^\infty dt \ t^{n-1 \over 2} \   e^{- (1-r)^2 \, t }
\sing  \frac{\Gamma({n+1 \over 2}) }{(1-r)^{n+1}}
\label{GX-4}
\end{equation}
as well as 
\begin{equation}
 \int_{t_0}^\infty dt \ t^{n-1 \over 2} \ \log t \  e^{- (1-r)^2 \, t }
 \sing 
 -2
 \frac{
 \Gamma({n+1 \over 2}) 
\log(1-r)}{(1-r)^{n+1}} + \frac{\Gamma'({n+1 \over 2}) }{(1-r)^{n+1}}
\label{GX-5}
\end{equation}
and
\begin{equation}
 \int_{t_0}^\infty dt \ t^{-1 } \   e^{- (1-r)^2 \, t }
\sing  - 2 \log(1-r)
\label{GX-6}
\end{equation}
one gets for  the two most singular parts of the right hand side of (\ref{BX-1})

\begin{align*}
 \int_0^\infty dt \ e^{r \nu_t - (1-r)^2 \, t}
 \sing  & 
\sum_{n \ge 0} \frac{b_n \, \epsilon^n  \, \Gamma\left({n+1 \over 2}\right)}{(1-r)^{n+1}} \\ &
+  \sum_{n \ge 1} \frac{c_n \, \epsilon^n  \,
\left( -2 \log(1-r) \Gamma\left({n \over 2}\right) + \Gamma'\left({n \over 2}\right)  \right)}
 {(1-r)^{n} }\\ &
  -  2 d_0 \log (1-r)  + \sum_{n \ge 1} \frac{d_n \, \epsilon^n \, \Gamma\left({n \over 2}\right)} { \ (1-r)^{n}} \\
& - (1-r)
\left[-{1 \over 2}   \sum_{n \ge 0}  \frac{b_n \,  \epsilon^n
\left( -2 \log(1-r) \Gamma\left({n+1 \over 2}\right) + \Gamma'\left({n+1 \over 2}\right)  \right)} {(1-r)^{n+1}} \right. \\
&  \ \ \ \ \ \ \left.+ \sum_{n \ge 0}\frac{  f_n \,  \epsilon^n  \Gamma\left({n+1 \over 2} \right)}{ (1-r)^{n+1}} \right]
\end{align*}
where the $f_n$'s are the coefficients of the series 
$$\sum_{n \ge0} f_n z^n = \left(\sum_{n \ge 0} b_n z^n \right) \ \log\left(\sum_{n \ge 0} b_n z^n \right)$$

Then equating the two most singular parts of the two sides  of (\ref{BX-1}) one gets 
This gives
\begin{align*}
b_n=& {1 \over \Gamma\left({n+1 \over 2}\right)}
 \ \ \ \ \   \text{for} \ \ \ n \ge 0 \\
c_n=& -{1 \over 2 \Gamma\left({n \over 2}\right)}
 \ \ \ \ \ \text{for} \ \ \ n \ge 1 \\
d_0=& -{1 \over 2 }   \\
d_n=&
{1 \over 2} {\Gamma'\left({n \over 2}\right) \over
\Gamma\left({n \over 2}\right)^2}
-{1 \over 2} {\Gamma'\left({n+1 \over 2}\right) \over
\Gamma\left({n \over 2}\right) \,
  \Gamma\left({n+1 \over 2}\right)}
-f_n  {\Gamma\left({n+1 \over 2}\right) \over   \Gamma\left({n \over 2}\right)}
-{n \over \Gamma\left({n \over 2}\right)}\ \ \ \ \  \text{for} \ \ \ n \ge 1 \\
\end{align*}
Except for the $d_n$'s  one can get closed expressions of the series (see Appendix A)
$$ \sum_{n\ge0} b_n z^n =  G_1(z) = 
 \sum_{n\ge0} {z^n \over \Gamma\left({n+1 \over 2}\right)}
$$
$$ \sum_{n\ge0} c_n z^n =G_0(z) 
=-{z \over 2} G_1(z) $$
This  gives  the dominant vanishing correction to (\ref{AX-17}) for the step initial condition (\ref{AX-2})
\begin{equation}
\mu_t =2 t -\frac{1}{2} \log t + \Psi_1( \epsilon \sqrt{t})  
 -\frac{1}{2} \log \pi - \frac{\epsilon \ \log t}{2 \sqrt{t}} + \frac{\sum_{n \ge 0} d_n (\epsilon \sqrt{t})^n}{\sqrt{t} \ G_1 (\epsilon \sqrt{t})}  + o\left(\frac{1}{\sqrt{t}} \right)
\label{GX-10}
\end{equation}
Given the values of $b_0$ and of $d_0$ it is easy to see that this is consitent with (\ref{AX-15}) when $\epsilon=0$ i.e. when $a=1$. On the other hand, I was not able  to estimate the series $d_n$ for $z\to -\infty$ and so could not check if (\ref{GX-10}) is consistent with (\ref{AX-15}) in the limit $\epsilon \to -\infty$.

% \newpage

\end{document}